\documentclass{emulateapj}

\def\spose#1{\hbox to 0pt{#1\hss}}
\def\simlt{\mathrel{\spose{\lower 3pt\hbox{$\mathchar"218$}}
     \raise 2.0pt\hbox{$\mathchar"13C$}}}
\def\simgt{\mathrel{\spose{\lower 3pt\hbox{$\mathchar"218$}}
     \raise 2.0pt\hbox{$\mathchar"13E$}}}

\slugcomment{AJ Accepted}

\shorttitle{Stellar Content of Leo~II}
\shortauthors{Komiyama et al.}

\begin{document}

\title{Wide-field Survey around Local Group Dwarf Spheroidal Galaxy Leo~II: 
	Spatial Distribution of Stellar Content$^{\!}$\altaffilmark{1}
}

\author{Yutaka Komiyama,$^{\!}$\altaffilmark{2,3}
	Mamoru Doi,$^{\!}$\altaffilmark{4}
	Hisanori Furusawa,$^{\!}$\altaffilmark{3} 
	Masaru Hamabe,$^{\!}$\altaffilmark{5}
	Katsumi Imi,
	Masahiko Kimura,$^{\!}$\altaffilmark{3}
	Satoshi Miyazaki,$^{\!}$\altaffilmark{3}
	Fumiaki Nakata,$^{\!}$\altaffilmark{2}
	Norio Okada,$^{\!}$\altaffilmark{2}
  	Sadanori Okamura,$^{\!}$\altaffilmark{6,7}
	Masami Ouchi,$^{\!}$\altaffilmark{8,9}
	Maki Sekiguchi,
	Kazuhiro Shimasaku,$^{\!}$\altaffilmark{6,7}
	Masafumi Yagi,$^{\!}$\altaffilmark{2}
	and Naoki Yasuda$^{\!}$\altaffilmark{10}
}

\altaffiltext{1}{Based on data collected at Subaru Telescope, 
	which is operated by the National Astronomical Observatory of Japan}
\altaffiltext{2}{National Astronomical Observatory of Japan, 
	2-21-1 Osawa, Mitaka, Tokyo 181-8588, Japan}
\altaffiltext{3}{Subaru Telescope, 650 North Aohoku Place, Hilo, 
	HI 96720, USA}
\altaffiltext{4}{Institute of Astronomy, Graduate School of Science,
	 University of Tokyo, 
	2-21-1 Osawa, Mitaka, Tokyo 181-0015, Japan}
\altaffiltext{5}{Department of Mathematical and Physical Sciences,
	Japan Women's University, Bunkyo, Tokyo 112-8681, Japan}
\altaffiltext{6}{Department of Astronomy, University of Tokyo,
	7-3-1 Hongo, Bunkyo, Tokyo 113-0033, Japan}
\altaffiltext{7}{Research Center for the Early Universe, School of Science,
	the University of Tokyo, Bunkyo, Tokyo 113-0033, Japan}
\altaffiltext{8}{Space Telescope Science Institute, 
	Baltimore, MD, USA}
\altaffiltext{9}{Hubble Fellow}
\altaffiltext{10}{Institute for Cosmic Ray Research, University of Tokyo, 
	Kashiwa 277-8582, Japan}

\begin{abstract}

We carried out a wide-field $V, I$ imaging survey 
of the Local Group dwarf spheroidal galaxy Leo~II
using the Subaru Prime Focus Camera on the 8.2-m Subaru Telescope. 
The survey covered an area of $26.67 \times 26.67$ arcmin$^{2}$,
far beyond the tidal radius of Leo~II (8.63 arcmin), 
down to the limiting magnitude of $V \simeq 26$, which is 
roughly 1 mag deeper than the turn-off point of the main sequence 
stars of Leo~II.  Radial number density profiles of bright 
and faint red giant branch (RGB) stars 
were found to change their slopes at around the tidal radius, and
extend beyond the tidal radius with shallower slopes.
A smoothed surface brightness map of Leo~II suggests the existence of 
a small substructure (4$\times$2.5 arcmin$^{2}$, 270$\times$170 pc$^{2}$
in physical size) of globular cluster luminosity beyond the tidal radius. 
We investigated the properties of the stellar population by means of
the color-magnitude diagram.
The horizontal branch (HB) morphology index shows a radial gradient
in which red HB stars are more concentrated than blue HB stars, 
which is common to many Local Group dwarf spheroidal galaxies.  
The color distribution of RGB stars around the mean RGB sequence
shows a larger dispersion at the center than in the outskirts,
indicating a mixture of stellar populations at the center
and a more homogeneous population in the outskirts.
Based on the age estimation using subgiant branch (SGB) stars, 
we found that although the major star formation took place 
$\sim$ 8 Gyr ago, a considerable stellar population 
younger than 8 Gyr is found at the center; such a younger
population is insignificant in the outskirts. 
The following star-formation history is suggested for Leo~II. 
Star-forming activity occurred more than
$\simgt$ 8 Gyr ago throughout the galaxy at a modest star-formation
rate. The star-forming region gradually shrank 
from the outside toward the center and star-forming activity finally 
dropped to $\sim$ 0 by $\sim$ 4 Gyr ago, except for 
the center, where a small population younger than 4 Gyr is present.

\end{abstract}



\section{Introduction}\label{sec:intro} 

Dwarf galaxies are the most numerous constituents in the universe 
and outnumber giant galaxies. 
In the prevailing hierarchical structure formation scenario 
(e.g., White \& Rees 1978; White \& Frenk 1991),  
they play key roles as building blocks from which 
larger structures such as giant galaxies are formed. 
Although most dwarf galaxies contain old stellar populations 
(Grebel 2000; Grebel \& Gallagher 2004), 
a general trend occurs in the evolution of galaxies 
in that more massive galaxies are formed at higher redshifts 
which is known as the "downsizing" effect  
(Cowie et al. 1996; Smail et al. 1998). 
It is therefore important to investigate how they evolved 
over the age of the universe. 

Extensive and epoch-making observations of Local Group dwarf galaxies 
using the Hubble Space Telescope (HST) markedly improved 
our knowledge of their evolutionary process. 
These observations have revealed their intriguing 
star-formation histories, which has been succinctly summarized as 
"no two Local Group dwarfs have the same star-formation history"
(Mateo 1998).  
However, the Achilles' heel of HST is its small field of view. 
In the Fornax dwarf spheroidal galaxy, 
Coleman et al. (2004, 2005) found "lobed" substructures, 
which are suggested to represent 
a disrupted merging companion dwarf galaxy 
located at $\sim 1.8$ core radii from the center
and outside the tidal radius.
Evidence for the existence of substructures is also suggested 
by both photometric and dynamical analyses for 
Ursa Minor (Kleyna et al. 1998; Wilkinson et al. 2004), 
Draco (Wilkinson et al. 2004), and 
Sextans (Kleyna et al. 2004; Walker et al. 2006). 
Extended halo structures are also found in 
several close companion dwarf spheroidals of the Milky Way 
(Ursa Minor, Palma et al. 2003; Carina, Majewski et al. 2005; 
Sculptor, Westfall et al. 2006), and their origin 
is often related to the tidal influence of the Milky Way. 
Regarding stellar populations, 
da Costa et al. (1996) first pointed out that 
Leo~II, And~I and Sculptor 
show a significant radial gradient in HB morphology. 
Since then many dwarf spheroidal galaxies have been reported to show  
radial gradients of stellar populations in the sense of 
a central concentration of young and metal-rich populations versus
more extended metal-poor and old populations 
(Mart\'inez-Delgado, Gallart \& Aparicio 1999; 
Saviane, Held \& Bertelli 2000; Harbeck et al. 2001; Tolstoy et al. 2004). 
However, some exceptions exist, such as Leo~I (Held et al. 2000) 
and Carina (Smecker-Hane et al. 1994; Harbeck et al 2001), 
although a mild radial gradient was reported for Carina (Koch et al. 2006). 
All these results demonstrate that even small dwarf galaxies, 
often described as simple systems, 
contain such complex structures inside. 
Hence, it is important to explore the whole galaxy 
from this perspective.
A combination of good image quality, depth, 
and a wide field of view is required for such purposes. 
One of the best facilities for conducting such observations 
is Suprime-Cam on the 8.2-m Subaru Telescope. 
We therefore carried out a wide-field imaging survey 
for the Local Group dwarf spheroidal galaxy Leo~II.

Leo~II is one of the Milky Way companion dwarf spheroidal galaxies 
located about 233 kpc from us (Bellazzini et al. 2005).
In contrast to the close companion dwarf spheroidal galaxies 
such as Sextans, Ursa Minor, and Draco, 
Leo~II resides in a relatively remote place from the Milky Way. 
The stellar content of Leo~II was studied extensively by 
Mighell \& Rich (1996) using WFPC2 on HST. 
They estimated the metallicity of Leo~II to be ${\rm [Fe/H]}=-1.60\pm0.25$ 
based on the $V, I$ color-magnitude diagram, which is consistent with 
a recent spectroscopic measurement by Bosler et al. (2004) 
who derived a median metallicity of ${\rm [Fe/H]}=-1.57$ 
based on the spectra obtained with Keck LRIS. 
They also noted that Leo~II started forming stars about 14$\pm$1 Gyr ago 
and formed most of its stellar population during the succeeding 7$\pm$1 Gyr, 
with a typical star having formed about 9$\pm$1 Gyr ago. 
A more recent study (Koch et al. 2007) showed that 
the mean metallicity of Leo~II is -1.74 based on the measurement of 
the calcium triplet for 52 red giants. 
These investigators also estimated individual ages, 
and derived a wide age range (2 - 15 Gyr, 
the same spread as found by Bosler et al. 2004) 
and an essentially flat age-metallicity relation. 
Dolphin (2002) reanalyzed the HST data and derived 
the star-formation history of Leo~II.  
He claimed that the mean metallicity (${\rm [Fe/H]}=-1.13$) is higher than 
the estimates of Mighell \& Rich (1996) and Bosler et al. (2004), 
owing to a young mean age of the stars in Leo~II (9.4 Gyr).  
However, the data are limited to the central small area 
(4.44 arcmin$^{2}$) within 
the core radius of the galaxy ($2'.9$, Mateo 1998). 
Recently, Bellazzini et al. (2005) published 
new $V, I$ photometry data obtained with the 3.5-m TNG 
covering a relatively wide area of Leo~II ($9.4\times9.4$ arcmin$^{2}$).  
They analyzed the spatial variation of the stellar content  
such as red clump stars and blue HB stars 
and the magnitude of the AGB bump, which indicates 
that the main population of Leo~II is $\simeq$ 8 Gyr. 
However, their data are shallow ($V_{lim}\sim22$) and 
their analysis inevitably limited to features brighter 
than the HB level. 
Our data obtained with Suprime-Cam on the 8.2-m Subaru Telescope 
constitute an excellent data set that gives a crucial clue 
for understanding the properties of the stellar content of Leo~II. 

In Section \ref{sec:obs}, we present the details of our observation 
and data analysis and show our results in 
Section \ref{sec:radprof} through \ref{sec:age}.
On the basis of these results, we discuss the formation and evolution 
of Leo~II in Section \ref{sec:evol} and give a summary 
in Section \ref{sec:summary}. 
Here we adopt the distance modulus of Leo~II
to be $(m-M)_{0}=21.63$ and the reddening to be $E(B-V)=0.02$ (Mateo 1998).  


\section{Observation and Data Analysis}\label{sec:obs}

The observation was carried out 
in April 2001 using the Subaru Prime Focus Camera
(Suprime-Cam; Miyazaki et al. 2002) on the 8.2-m Subaru Telescope 
at Mauna Kea, Hawaii.
Suprime-Cam is a wide-field imager consisting of 10 2k$\times$4k CCDs. 
It covers a sky area of $34\times27$ arcmin$^{2}$ 
with 0.2 arcsec per pixel sampling.
Because of the wide-field coverage and good image quality of 
the Subaru Telescope, 
Suprime-Cam is the most powerful instrument for investigating
stellar contents of nearby galaxies.
We used $V$ and $I$ filters and total exposure times
are 3000 sec and 2400 sec in the $V$ and $I$ bands, respectively.
Several short exposures were also obtained to measure the luminosities 
of bright stars, which are saturated in long exposure frames.
The sky condition was good and the typical stellar size (FWHM)
was about 0.7 arcsec in both $V$ and $I$ bands.
The details of the observation are given in Tab.~\ref{tab:obs}.

The data were reduced using the standard data analysis software
for Suprime-Cam (Yagi et al. 2002). 
The reduction procedure is summarized as follows.
The bias was subtracted from individual frames and bad pixels were masked.
Each frame was divided by the flat frame, which was created from
object frames (mostly blank fields) taken in the same observing run.
Note that Leo~II frames were excluded when creating the flat frames.
The optical distortion caused by the prime focus corrector
was corrected using an analytical expression of the optical distortion
(see Miyazaki et al. 2002), and the sky background was subtracted
from each frame. Then the two-dimensional position and the relative
brightness of each frame were estimated using common stars
found in adjacent CCD chips and different dither positions.
Finally we obtained a coadded image.
The FWHMs of stars in the resultant coadded images are
0.80 arcsec and 0.78 arcsec in the $V$ and $I$ bands, respectively.
We used the central area (8000$\times$8000 pixels,
26.67$\times$26.67 arcmin$^{2}$) for the following analysis to guarantee
a homogeneous signal-to-noise ratio over the wide field of view.
As shown in Fig.~\ref{fig:leo2}, the survey area is wide enough
and far beyond the tidal radius of Leo~II (8.7 arcmin; Mateo 1998).

We applied DAOPHOT PSF photometry software (Stetson 1987, 1994)
for the coadded images.
The PSF model was made from about 100 stars and we repeated iterations
of the PSF photometry three times to not miss faint stars.
Non-stellar objects such as galaxies and cosmic rays were
excluded using shape and $\chi^{2}$ parameters calculated by DAOPHOT.
Combining bright stellar objects ($V<20$) detected in short exposure frames 
and faint stellar objects ($V>20$) detected in long exposure frames, 
82252 objects were cataloged as stellar objects.

Zero-point magnitudes in both $V$ and $I$ bands were calibrated
using bright stars ($V<20$) listed in Lee (1995).
We used short exposures for comparison since the bright stars
listed in Lee (1995) were saturated in long exposure frames.
The zero-point magnitudes are accurate to 0.03 mag and 0.01 mag
in the $V$ and $I$ bands, respectively.
Long exposure frames were calibrated using common stars
(typically 20-22 mag stars) on both long and short exposure frames.
Long exposure frames are accurate to 0.01 mag (relative to short exposures)
in both bands.

The magnitude error and the detection completeness were estimated
in the standard manner. 
We divided the 8000$\times$8000 pixel image into
80 $\times$ 80 grids consisting of 100$\times$100 pixels. 
In each grid, an artificial star was added at random position
using the {\tt addstar} task in the DAOPHOT package, 
and the same PSF photometry procedure was applied to the image. 
This process was repeated for 10 times per every
0.5 magnitude interval for the magnitude ranges of 
23.5 mag $<V<$ 26.0 mag and 22.5 mag $<I<$ 25.0 mag, respectively. 
The magnitude error and the detection completeness were calculated 
from the result of the PSF photometry for these artificial stars. 
The result for the $V$ band is shown in Fig.~\ref{fig:complete} 
as a function of magnitude and the distance from the galaxy center.
The detection completeness is $>0.9$ for $V<24.5$ at any position 
in the galaxy, but it degrades to 0.6 at the galaxy center for $V=25.5$. 
The 90\% and 50\% completeness limits at the galaxy center 
are 24.5 and 25.9 in $V$ band, respectively, 
and those for $I$ band are 22.7 and 24.7, respectively.
The magnitude is accurate to 0.02 mag for $V<24.5$ in most parts of 
the galaxy, but the degradation is severe at the galaxy center. 
For $V<23.5$ and $I<22.5$, the detection is almost complete and 
the magnitude is accurate even at the crowded galaxy center. 

Fig.~\ref{fig:cmd} shows the color-magnitude diagram of stellar objects 
found in the central $6.67\times6.67$ arcmin$^{2}$ area of the Leo~II field.
It is clearly seen that our data cover a wide magnitude range
of stars in Leo~II from the tip of the RGB ($V\simeq19$)
to the turn-off point ($V\simeq25$). 
Remarkable features are the well-defined HB
at $V\simeq22.2$ and the narrow RGB. 
The red HB is characterized by a concentration of stars 
at the red side of the RR Lyr instability strip ($0.4 \simlt V-I \simlt 0.6$)  
that is well distinguished from the RGB. 
The HB extends to the blue side and forms another concentration 
at $0 \simlt V-I \simlt 0.4$. 
It is obvious that the asymptotic giant branch (AGB) merges 
into the RGB at $V\sim21.5$ and the 
RGB bumps detected by Bellazzini et al. (2005) 
are clearly seen by eye at $V \sim 21.4$ and $V \sim 21.8$.
One might notice that $\sim$20 stars with the same color as 
the red HB, but much brighter than the red HB, occur. 
They may possibly be helium-burning, high-mass stars 
(Mighell \& Rich, 1996; Bellazzini et al. 2005),  
although Demers \& Irwin (1993) first argued that they are 
a photometric blend of HB and RGB stars. 
The other noteworthy feature in the color-magnitude diagram is 
the apparent bifurcation of the blue HB stars. 
The feature is also seen in Bellazzini et al. (2005; Fig.2),
and according to their identification, most of the brighter 
blue HB stars are variable stars cataloged by Siegel \& Majewski (2000).
We examined the spatial distribution of these stars and 
found no particular maldistribution (concentrated or uniform distribution).

We note that the contamination from Galactic stars
is not severe compared to other Milky Way satellite galaxies
(e.g., Sextans, Draco, and Ursa Minor; See Harbeck et al. 2001)
since Leo~II is located at a relatively 
high galactic latitude ($b=67^{\circ}$).
The contamination becomes severe for $V>23.5$.
The typical photometric errors, which were calculated on the basis 
of the artificial star test (thus including the effect of the crowding), 
are plotted as blue (near center) and red (outskirts) error bars 
in Fig.~\ref{fig:cmfea}~(a).


\section{Radial Distribution of the Stellar Component}\label{sec:radprof}

We first investigated the radial profiles of 
bright and faint RGB stars and blue and red HB stars. 
The blue and red HB stars are 
easily discerned as seen in Fig.~\ref{fig:cmd}. 
We defined the blue HB stars as $0<V-I<0.38, 21.88<V<22.48$ stars and 
the red HB stars as $0.58<V-I<0.88, 21.88<V<22.38$ and 
$V> -0.4/0.16 [(V-I)-0.58] +22.08$. 
See Fig.~\ref{fig:cmfea} for these criteria in detail. 
To identify RGB stars, 
we determined the mean RGB sequence which was fitted as, 
\begin{equation}
  (V-I)_{RGB} = 197.717 - 33.592 V + 2.169 V^{2}  
	- 6.267\times 10^{-2} V^{3} + 6.830\times 10^{-4} V^{4}
\label{eq:rgbseq}
\end{equation}
Fig.~\ref{fig:cmfea}~(a) shows how well the mean RGB sequence traces the data. 
The stars that deviate less than $\pm$0.075 mag 
(corresponding to 2.3$\sigma$) in $V-I$ color 
from the mean RGB sequence are classified as RGB stars. 
The criteria enclose most of the RGB stars and separate 
red HB stars fairly well. 
We set the faint limit of the RGB at $V=23.5$ 
to avoid contamination 
from foreground stars and unresolved background galaxies, 
as well as to be free from the completeness correction. 
The RGB stars were subdivided into bright and faint RGB stars 
at the HB level ($V_{HB}=22.18$, Mighell \& Rich 1996).  

We compared the mean RGB sequence with those of Galactic 
globular clusters M~15, NGC~6397, M~2, and NGC~1851
taken from da Costa \& Armandroff (1990) in Fig.~\ref{fig:cmfea}~(b). 
These clusters have metallicities [Fe/H] of -2.17, -1.91, -1.58, and -1.29, 
respectively (da Costa \& Armandroff 1990). 
The mean RGB sequence of Leo~II lies in between NGC~6397 and M~2,  
suggesting that the mean metallicity of Leo~II would be 
between -1.91 and -1.58 
if an old stellar population as Galactic globular clusters is assumed. 
This value is consistent with those 
derived  spectroscopically by Bosler et al. (2004) and Koch et al. (2007).   
The mean RGB sequence we obtained is slightly bluer than 
that derived by Mighell \& Rich (1996). 
Their mean RGB sequence is just on the M~2 RGB sequence. 
A likely cause of this could be the difference
in the size of the survey field and will be discussed further 
in Sect.~\ref{sec:rgb}.  

We counted the number of stars in each stellar component 
(i.e., bright and faint RGB, blue and red HB) in an annular area 
of $r_{in}<r<r_{out}$ and divided this by the area of the annulus 
to derive the number density. 
The characteristic radius $<r>$ for an annulus is defined as, 
\begin{eqnarray}
  \int_{r_{in}}^{<r>} dA &=& \int_{<r>}^{r_{out}} dA \\
  \langle r \rangle
      &=& \sqrt{(r_{out}^{2}+r_{in}^{2})/2}
\end{eqnarray}
In Fig.~\ref{fig:radprof} the radial profiles for each stellar component 
are plotted as a function of the characteristic radius. 
The numbers are listed in Tab.~\ref{tab:radprof}.  
We fitted the radial profile for each stellar component with the 
King profile and listed the best-fit parameters in Tab.~\ref{tab:king}. 
The core and tidal radii calculated for all RGB stars are 
2.76 arcmin and 8.63 arcmin, respectively, 
and are consistent with those derived by Irwin \& Hatzidimitriou (1995). 
Bright RGB stars are slightly more concentrated than faint RGB stars 
in terms of the core radius. 
This is also confirmed by a cumulative number fraction plot 
shown in the inset of Fig.~\ref{fig:radprof}. 
We calculated the KS probabilities that two different stellar components 
had the same spatial distribution. The probabilities are less than 1\% 
except for the pair of bright RGB and red HB stars (76.3\%). 
The King profile fitting for bright RGB stars is achieved for $r<9$ arcmin, 
as suggested by the best-fit tidal radius of 9.22 arcmin, 
and the number density of bright RGB stars 
shows the shallower slope for $r>9$ arcmin 
and a probable drop at $r>14$ arcmin.  
A similar trend is also seen for faint RGB stars, and 
the change in the slope occurs at $r\sim 8.5$ arcmin 
(c.f., the best-fit tidal radius of 8.51 arcmin), 
although the number density may reach the field level 
at $r>11$ arcmin. 

The field contamination is estimated in the following way. 
Ratnatunga \& Bahcall (1985) calculated the number of 
field stars with different colors toward the direction of Leo~II. 
The number of field stars with $(B-V)<0.8$ and $19<V<22.18$ 
is estimated to be 0.14 arcmin$^{-2}$ based on their table. 
Considering that the color of $(B-V)=0.8$ corresponds to a K0V star and 
hence, $(V-I)=1.87$, and that most field stars are redder than $(V-I)=0.6$, 
we expect 0.14 arcmin$^{-2}$ field stars 
in the color range of $0.6<(V-I)<1.87$. 
We therefore estimated that 0.0165 arcmin$^{-2}$ 
field stars are in our bright RGB selection criteria
($19<V<22.18$ and $\Delta(V-I)=0.15$). 
We also estimated the number of field stars using 
the SDSS DR5 archive data (Adelman-McCarthy et al. 2007). 
The bright RGB selection criteria were determined on 
the basis of the (g, i) color-magnitude diagram of Leo~II 
and the number of stars within the criteria 
in the nearby field of Leo~II (1 degree from Leo~II) was determined.
The estimated field contamination is 0.0226 arcmin$^{-2}$, 
which is consistent with that determined above. 
We therefore conclude that the number of field contaminations 
for the bright RGB stars is $\sim 0.02$ arcmin$^{-2}$
and that stars located at $r>14$ arcmin are likely to 
be dominated by the field population. 
Adopting this field contamination number, 
we suggest that the shallower slope of the radial profile found 
for $9<r<13$ arcmin is real. 
The field contamination for faint RGB stars is expected 
to be smaller than $\sim 0.02$ arcmin$^{-2}$ 
because of the smaller magnitude coverage of the selection criteria, 
but contamination from background compact galaxies 
that are misclassified as stars may occur. 
The SDSS data are too shallow to be used for 
estimating the field contamination. 
If stars found for $r>14$ arcmin consist of such a mixture of 
field contamination and the background compact galaxies 
as implied from the analysis for the bright RGB stars, 
the shallower slope found for $8<r<11$ arcmin 
is also suggested to be real. 


To further investigate the details of the extra-tidal structure, 
we made a smoothed surface brightness map for the entire survey field 
as follows. Stars regarded as RGB or HB stars were listed 
and Gaussians of 1 arcmin kernel multiplied by 
the luminosity of each star was placed at the position of each star. 
They were then coadded to obtain a smoothed surface brightness map. 
This operation makes hidden faint structures clearer. 
Fig.~\ref{fig:densmap} is the resuling smoothed surface brightness map.  
The isosurface-brightness contour of the bright part of 
the galaxy is almost circular, but it becomes more complicated 
at a lower surface brightness.  
The most remarkable feature of Fig.~\ref{fig:densmap} is the diffuse 
knotty structure prominent in the eastern part of the galaxy 
($\Delta\alpha \sim$ = -11, $\Delta\delta \sim$ 1). 
The knot is more than five times more luminous than 
the position located at the same distance from the center 
at the opposite side of the galaxy, 
although the mean surface brightness is quite faint 
($\sim$ 31 mag/arcsec$^{-2}$). 
The knot contains four bright RGB stars in $\simeq 4 \times 5$ 
arcmin$^{2}$ area and the expected field contamination number is 0.4, 
indicating that the knot is 99.92\% significant above the field population 
on the basis of Poisson statistics. 


The extent of this knot is about 4 arcmin (270 pc in physical size) 
with a width of 2.5 arcmin (170 pc), 
and it is small compared to the main body of Leo~II. 
The magnitude of this knot was estimated to be $M_{V}=-2.8$  
by summing up luminosities of 15 stars found in the knot region 
that are brighter than $V=23.5$. 
The value is close to the magnitude of the least luminous globular cluster. 
The knot must be more luminous because we neglected
a contribution from underlying faint stars, and could be more luminous
if it is indeed extended farther to the east (out of our survey field),
or if the main part of it is already merged with the main body of Leo~II.
It is possible that the substructure is
a small globular cluster that is being disrupted
and merging into the main body of Leo~II. 
The other possibility is that the knot is composed of stars  
stripped from the main body of Leo~II.
The origin of the substructure is discussed further in Sect.~\ref{sec:evol}.


\section{Horizontal Branch Morphology}\label{sec:hbmorph}

In brief, the HB morphology indicates a distribution in the color of 
HB stars. It is often parameterized as $(B-R)/(B+V+R)$, 
where $B$ and $R$ are the numbers of 
blue and red HB stars, respectively, and $V$ is the number of stars 
lying on the RR Lyr instability strip. 
Intensive investigation on the HB morphology of globular clusters 
has shown that it depends primarily on metallicity 
in that less metal-rich systems show a bluer HB morphology, 
but it is also influenced by the {\it second parameter}, which is 
most likely to be age (Lee, Demarque \& Zinn, 1994).
The HB morphology is thus
a key measure in studying the properties of stellar populations 
and the variation in the HB morphology within a galaxy 
is often investigated (e.g., Harbeck et al. 2001; Tolstoy et al. 2004). 
Using our data, we can examine the detailed variation of 
the HB morphology over a wide radius 
from the center to far beyond the tidal radius of Leo~II. 

Fig.~\ref{fig:hbmorph} shows the HB morphology index 
$(B-R)/(B+V+R)$ plotted as a function of the radius. 
The index is less than zero at any radius, 
indicating that red HB stars are more numerous than blue HB stars
everywhere in Leo~II. 
This value agrees with those obtained in other studies 
(-0.68, Demers \& Irwin 1993; $-0.78\pm0.10$, Mighell \& Rich 1996). 
The index is small at the center of the galaxy and 
becomes larger as the radius increases for $r>3$ arcmin, 
reaching its maximum at $r=6$ arcmin. 
The trend is consistent with the findings of da Costa et al. (1996). 
They showed that the HB morphology index is approximately constant 
out to $r \simeq 3$ arcmin but the fraction of blue HB stars 
increases beyond $r \simeq 3$ arcmin. 
This means that red HB stars are more concentrated 
to the center than blue HB stars for $r<6$ arcmin.
The inset of Fig.~\ref{fig:hbmorph}, which presents 
the cumulative number fraction of blue and red HB stars 
as a function of the radius, clearly shows this 
and confirms the result of Bellazzini et al. (2005; see their Fig. 8). 
They suggest that age is the main driver of the population gradient. 
Koch et al. (2007) support this suggestion 
although they did not detect any considerable metallicity 
or age gradient in Leo II.
The trend of a centrally-concentrated red HB distribution
is also observed in many dwarf spheroidal galaxies 
in the Local Group 
(Majewski et al. 1999; Harbeck et al. 2001; Tolstoy et al. 2004). 
Our results support the idea that  
the radial gradient of the HB morphology is common to 
dwarf spheroidal galaxies. 

For the outer part of the galaxy ($r>7$ arcmin), 
the HB morphology index looks almost constant 
at $(B-R)/(B+V+R) \sim -0.6$, 
and the value is larger than that at the inner part ($r<5$ arcmin). 
This means that blue HB stars are more numerous, 
implying that the stellar population in the outer region
is less metal-rich and/or older than
those in the inner part.


\section{Blue/Red RGB Distribution}\label{sec:rgb}

We investigated the color distribution of the RGB stars. 
In an analogy to the HB morphology index, 
we used the RGB color index for the analysis, defined as
$(B-R)/(B+R)$, where $B$ and $R$ are the numbers of stars that 
deviate less than 0.075 mag bluer and redder from the mean RGB sequence,
respectively (see also Fig.~\ref{fig:cmfea}).
The mean RGB sequence is defined as Eq.~\ref{eq:rgbseq}, and 
those stars $19<V<23.5$ were used. 
Since the AGB merges to the RGB from the blue side to 
to the bright part of the RGB, 
it is possible that the RGB color index may not have been determined correctly
due to the contamination of AGB stars, especially when 
the number fraction of AGB stars to RGB stars is large.  
To estimate the influence of AGB stars in the determination of the index, 
we derived the RGB color index using whole RGB stars 
($19<V<23.5$) and faint RGB stars ($22.18<V<23.5$).  
We plotted the results as open triangles (whole RGB) and 
filled squares (faint RGB) in Fig.~\ref{fig:rgbmorph}.  
The color index derived from whole RGB stars at a fixed radius 
is slightly larger (i.e., bluer color) than 
that derived from faint RGB stars,  
indicating an influence, albeit small, of AGB stars. 
Therefore, the RGB color index is more accurately derived 
by using faint RGB stars ($22.18<V<23.5$). 

The color index is distributed around zero at any radius
except for the center where red RGB stars seem to be numerous. 
This fact gives a reasonable explanation for 
the color difference of the mean RGB sequence
between this study and the redder mean RGB color of Mighell \& Rich (1996). 
Since their survey was limited to a small area (4.44 arcmin$^{2}$) 
at the galaxy center, 
they inevitably sampled red RGB stars, which are numerous at the center, 
and hence obtained a redder mean RGB color. 
This also suggests that the stellar population varies 
within a galaxy. 

The inset of Fig.~\ref{fig:rgbmorph} shows the 
cumulative number fraction of both blue and red RGB stars. 
The radial distribution is quite similar between the blue and red RGB stars,  
in contrast to the same figure for blue and red HB stars 
(Fig.~\ref{fig:hbmorph}). 
However, the coincidence of the RGB color indices of 
the two stellar groups does not always mean that the stellar populations 
of two groups are identical. For example,  
the color index cannot distinguish between 
broad and narrow color distributions around the mean RGB sequence; 
thus, examining the color distributions around this sequence 
is of key importance, as shown in Fig.~\ref{fig:frgbchist}. 
Here we divided the stars into four groups according to radius, 
$r<1'.5$, $1'.5<r<3'.0$, $3'.0<r<6'.7$ and $6'.7<r$, 
and made a color histogram for each group. 
The figure shows that the color distribution is generally broad, 
but varies as the radius changes. 
It appears that the color distribution for $r<3'.0$ is 
very broad, suggesting that 
the stellar population at the galaxy center is not simple
and is a mixture of several stellar populations of 
different ages and metal abundance. 
This is consistent with the results of 
Mighell \& Rich (1996), who noted the wide age 
spread ($\sim 7$Gyr) for the stellar population at the center. 
The color distribution becomes 
more concentrated to $\Delta (V-I)=0$ for $r>3.0$ arcmin. 
This would imply that the stellar population for $r>3.0$ arcmin 
is more homogeneous compared to that for $r<3.0$ arcmin.
\footnote{Note that a narrow color distribution does not necessarily imply 
a homogeneous stellar population (e.g., Smecher-Hane et al. 1994)}.


\section{Radial Gradient of Age Distribution}\label{sec:age}

Mighell \& Rich (1996) derived the age distribution of 
the stellar population in the center of the galaxy 
on the basis of the magnitude distribution of subgiant branch (SGB) stars. 
Fig.~\ref{fig:cmfea}~(c) focuses on the color-magnitude diagram 
around the bottom of the RGB and the turn-off point. 
The green lines represent Padova isochrones for ages 5, 10, and 15 Gyr 
and metallicity Z=0.0004 (Girardi et al. 2002). 
As shown in the figure, the isochrones are almost parallel 
to the $V-I$ axis (i.e., constant $V$ magnitude) 
at the SGB ($V-I \simeq 0.7$), indicating that the magnitude 
at a fixed $V-I$ color can be translated to age. 
The difference in metallicity also affects the shape of the isochrone, 
but small differences in metallicity 
(e.g. Z=0.001, shown as magenta lines in Fig.~\ref{fig:cmfea}~(c)) 
do not change the shape significantly. 
Since it is unlikely that metal rich population ($Z>0.004$) 
dominates the stellar population in Leo~II 
as suggested by the shape of mean RGB sequence, 
we can estimate the age distribution 
using the magnitude distribution of SGB stars. 
We examined the magnitude distribution of stars 
with $23.5<V<25.5$ and $0.67<V-I<0.77$, which we call SGB. 
The region in the color-magnitude diagram is shown 
as a cyan box in Fig.~\ref{fig:cmfea} (a) and (c).   

A difficulty, however, occurs in applying this method to our data; 
the error in color becomes larger than 
the width of the selection criteria, $0.67<V-I<0.77$, for $V>24.5$. 
This increases the uncertainty in the number estimation of 
SGB stars fainter than $V=24.5$. 
Nevertheless, we were able to obtain a clue as to
the age distribution in the following way. 
The key lies in the brighter SGB stars ($V<24.5$), which 
indicate the presence of a younger stellar population. 
We can estimate what fraction of the total stellar population 
the young population accounts for 
by comparing the number ratio of bright SGB stars to faint RGB stars 
with a theoretical calculation. 
We therefore investigated the number ratio of SGB stars to faint RGB stars 
as a function of the radius.  

To derive the number of SGB stars, the incompleteness of the detection and 
contaminations from unresolved background 
galaxies and the foreground stars must be properly corrected. 
We estimated the incompleteness for every 
0.5 mag grid in the $V$ and $I$ bands 
and for 1 arcmin in radius  
using real images as explained in Sect.~\ref{sec:obs}. 
Fig.~\ref{fig:comp_mag} shows the completeness as a function of 
magnitude in the $V$ and $I$ bands at different radii 
($r$ = 0, 2.5, 5.0, 10.0 arcmin). 
With this incompleteness table in hand, 
the incompleteness at a given $V$ magnitude, color 
(i.e., $I$ magnitude, once $V$ magnitude is given), and radius 
is estimated by a linear interpolation. 
The numbers of SGB stars are corrected for incompleteness 
calculated above. 
To estimate the number of contaminations, 
we regarded stars found at $r>16.67'$ as contaminations, 
and the magnitude distribution of (incompleteness-corrected) 
contaminations with $0.67<V-I<0.77$ 
were fitted to the 4th order polynomials as, 
$C (arcmin^{-2} / 0.1 mag) = -33245 + 5448.7 V - 334.59 V^{2} 
	+ 9.1314 V^{3} - 0.093365 V^{4}$.
To derive the number of SGB stars in a given annulus, 
the contamination number function $C$ multiplied by the area 
of the annulus wa subtracted from the incompleteness-corrected number. 

The number ratios of SGB stars to faint RGB stars 
are plotted in Fig.~\ref{fig:sgfrgbratio} as a function of the radius. 
In the figure, the number ratios are plotted separately 
for bright SGB stars ($23.5<V<24.0$, filled squares) and 
intermediate SGB stars ($24.0<V<24.5$, open triangles). 
Note that $23.5<V<24.0$ and $24.0<V<24.5$ populations 
roughly correspond to $2.5 \sim 4$ Gyr and $4 \sim 6.3-8$ Gyr 
populations, respectively. 
We noted that the number ratios for both bright and 
intermediate SGB stars increase toward the center of the galaxy. 
The slope is steeper for intermediate SGB stars. 

The number ratio can be calculated theoretically 
for a stellar population of fixed age and metallicity 
using Padova isochrones and the initial mass function.  
We adopted Salpeter's form for the initial mass function. 
The calculation shows that the number ratios for bright SGB stars 
($23.5<V<24.0$) range $0.37 \sim 0.41$ for Z=0.0004 population stars. 
If a stellar population is dominated by a Z=0.0004 population, 
the number ratio should be close to the value. 
The number for a Z=0.001 population ranges $0.66 \sim 0.76$. 
Although the calculated values are different according to 
the adopted metallicity, 
the number ratios at any radius are well below all the calculated values.
This indicates that a population younger than 4 Gyr
is not a dominant population, although it 
certainly resides in the galaxy center. 
The existence of a stellar population as young as 2 Gyr 
reported by Bosler et al. (2004) and Koch et al. (2007) 
also supports our finding. 
The increase in the number ratio at the galaxy center suggests that 
(1) the fraction of the young population is higher at the center 
than at the periphery, 
(2) the metallicity of the young population is higher at the center 
than at the periphery, or 
(3) a combination of (1) and (2). 

For intermediate SGB stars ($24.0<V<24.5$), the calculated number ratios
range $0.5 \sim 0.8$ and $0.6 \sim 1.0$ for 
Z=0.0004 and Z=0.001 populations, respectively. 
The number ratio is $\sim 0.7$ at the center and $\sim 0.5$ 
within 3 arcmin from the center, indicating that an 
intermediate age population ($4 \sim 8$ Gyr) is dominant 
at the galaxy center. 
This is consistent with the finding by Mighell \& Rich (1996)
and Dolphin (2002) that a considerable stellar population 
younger than 8 Gyr occurs at the center of Leo~II.  
However, the number ratios of both bright and 
intermediate SGB stars become small as the radius increases, 
indicating that the stellar population at the outskirts of the galaxy 
is deficient in young population, i.e., most of the stars 
are older than 8 Gyr.


\section{The Evolution of Leo~II}\label{sec:evol}

\subsection{Main Body}

The stellar population in the outskirts of the galaxy 
($5 \simlt r \simlt r_{t}$) was shown to consist of  
mostly older stars ($\simgt$ 8 Gyr). 
If metal abundance is nearly homogeneous,  
such an old population must form a narrow color distribution at the RGB,  
which is confirmed by a concentrated distribution in $V-I$ color of 
faint RGB stars as shown in Fig.~\ref{fig:frgbchist}. 
A comparison of Padova isochrones with the color distribution of 
RGB stars in the outskirts suggests low-metal-abundance populations 
(between Z=0.0004 and Z=0.001) in the outskirts 
if ages of $10 \sim 15$ Gyr are assumed. 
The larger HB morphology index (Fig.~\ref{fig:hbmorph}) also 
supports an old population with low metal abundance. 
We conclude that the dominant population in the outskirts of the galaxy 
is an old population with low metal abundance.  

The stellar population at the center of the galaxy, however,  
shows a variety of age.  
It is necessary to include stars younger than 10 Gyr,  
but a young population with low metal abundance, for example,  
$\simlt$ 10 Gyr and Z=0.0004 population, is excluded 
since the isochrone would not trace the RGB distribution. 
Therefore, a higher metal abundance (Z $\simeq$ 0.001, 
possibly Z $\simeq$ 0.004 for very young population) is suggested. 

From the foregoing results, Leo~II is suggested to have evolved as follows. 
Leo~II first started to form stars over the whole galaxy
about 15 Gyr ago\footnote{This estimate is based on the oldest ages 
in the adopted isochrone grids.} 
with a modest (probably low) star-formation rate.
Star formation lasted for some time and the interstellar
gas gained metals.
Then about 8 Gyr ago, star formation began to cease from the
outskirts and the star-forming region gradually became 
more concentrated to the center.
The star-forming activity had dropped to $\sim$ 0 by $\sim$ 4 Gyr ago, 
except for the center where a small population younger than 4 Gyr occurs.

Hensler et al. (2004) demonstrated the one-dimensional 
chemodynamical evolution of dwarf elliptical galaxies, 
and showed the interesting feature that 
the star-forming region shrinks as a galaxy evolves 
because of gas exhaustion in the galaxy. 
Their simulation seems to outline the evolution of Leo~II fairly well, 
although it requires a refinement to fully explain our results. 
Since a population gradient within a galaxy is also observed for 
several Local Group dwarf spheroidal galaxies (e.g., Harbeck et al. 2001),  
a more refined chemodynamical model 
will be necessary to explain the population gradient 
in the future to clarify the evolution of dwarf spheroidal galaxies.

\subsection{Halo Structure}

The origin of the knotty substructure found at 
the extended halo of Leo~II could be 
(1) a small globular cluster, which is disrupted 
and merged with the main body of Leo~II, 
(2) stars stripped from the main body of Leo~II, or
(3) a foreground artifact. 
The properties of stellar populations such as HB morphology
are almost the same between outside the tidal radius 
and at the outskirts of the main body, indicating that 
the knot would be dominated by old stars with low metal abundance. 
To further investigate the stellar population of the knot, 
we made a Hess diagram from which field contaminations were 
statistically subtracted. 
In Fig.~\ref{fig:hess}, although the field subtraction is not perfect, 
two significant concentrations of stars are observed 
around the red clump ($V-I \sim 0.8$, $V\sim 22$) 
and the turn-off point ($V-I \sim 0.7$, $V \sim 26$) 
like that seen in Fig.~\ref{fig:cmd}. 
This suggests that the knot is likely to consist of a 
similar stellar population as that residing in the outskirts of Leo~II
and the probability of (3) is low. 
However, based on this figure, it is still difficult to determine 
whether possibility (1) or (2) is more likely. 


If the second scenario is true, 
the tidal influence of the Galaxy would be the most efficient 
mechanism to strip stars from the main body of Leo~II. 
Indeed, many dwarf spheroidal galaxies such as Draco and Ursa Minor 
are now known to host extra-tidal halo structures 
although they are closer to the Galaxy and hence 
more influenced by the Galactic tidal force. 
However, the present-day remote location of Leo~II from the Galaxy 
raises the question of whether the tidal force of the Galaxy is enough to 
strip stars from the main body of Leo~II. 
In addition, the fact that we do not detect any obvious 
extra-tidal structure at the opposite side of Leo~II 
is unfavorable for this scenario. 
Therefore, it is unlikely that tidally stripped stars are 
the origin of the knotty substructure. 
If the knot is indeed a result of the tidal stripping, 
it should be aligned to the direction parallel to the motion of Leo~II. 
Therefore, measuring the proper motion of Leo~II 
would provide a clue to answering this problem, 
although it would still be quite challenging. 


The fact that no globular clusters are found to associate with 
less luminous dwarf spheroidals such as Leo~II 
does not support the first scenario for the origin of the knot. 
But it is possible 
that Leo~II formed together with a small number of globular clusters 
and we may be watching the disruption process of the 
last one that happened to survive until the recent past. 
It is interesting that Kleyna et al. (2003) demonstrated 
the survival of a substructure for a Hubble time 
in a cored dark-matter halo. 
They suggested that the substructures found in Ursa Minor 
are the remnants of a disrupted stellar cluster and 
that Ursa Minor possesses a cored dark-matter halo. 
Following their idea, we suggest that 
Leo~II may be another example of a galaxy with a cored dark-matter halo. 

Recent numerical simulations suggest that 
dark halos of dwarf spheroidals are larger than previously 
thought, and hence, extra-tidal stars are 
gravitationally bound to the galaxies and 
are a part of the extended stellar halos 
(Hayashi et al. 2003; Mashchenko et al. 2005). 
The extended halo structure found in this study 
might be a structure bound to Leo~II 
according to the predictions of the simulations.


\section{Summary}\label{sec:summary}

We carried out a wide-field imaging survey of 
the Local Group dwarf spheroidal galaxy Leo~II in the $V$ and $I$ bands
using Suprime-Cam on the 8.2-m Subaru Telescope. 
The survey covered an area of $26.67 \times 26.67$ arcmin$^{2}$,
far beyond the tidal radius of the Leo~II (8.63 arcmin). 
A total of 82252 stars was detected down to 
the limiting magnitude of $V \simeq 26$, which is roughly 
1 mag deeper than the turn-off point of the main sequence stars of Leo~II. 
Our main conclusions are summarized below. 

\begin{itemize} 

\item 
The radial number density profile of bright RGB stars 
is shown to continue beyond the tidal radius ($r_{t} = 9.22$ arcmin).  
A change in the slope occurs near the tidal radius 
and the slope becomes  shallower for $r>9$ arcmin. 
A hint of a drop is seen in number density at $r>14$ arcmin. 
A similar two-component profile is also observed for faint RGB stars. 

\item
A smoothed surface brightness map of Leo~II suggests the existence of 
a small substructure beyond the tidal radius, 
which is as large as globular clusters in luminosity ($M_{V}<-2.8$). 
It could possibly be a disrupted globular cluster of Leo~II 
that had survived until the recent past. 
Another possibility is that it is composed of stars stripped 
from the main body of Leo~II, although this is unlikely.

\item
The HB morphology index shows a radial gradient  
in the sense that red HB stars are more concentrated than blue HB stars.  
Such a trend is also observed in several 
Local Group dwarf spheroidal galaxies. 
The HB morphology index implies that the stellar population in the outer part 
($r>7$ arcmin) is more metal-poor and/or older  
than that in the inner part.

\item
The RGB color index is almost constant at any radius 
except for the center, where a redder mean RGB sequence than ours 
was observed by Mighell \& Rich (1996). 
The color distribution of RGB stars around the mean RGB sequence 
shows a broader distribution at the center ($r<3$ arcmin) 
than the outskirts.  
This suggests a more homogeneous stellar population at the outskirts 
of the galaxy and a variety of stellar populations at the galaxy center. 

\item
The age distribution was estimated using brighter 
($23.5<V<24.5$) SGB stars. 
The presence of a younger stellar population than 4 Gyr is 
suggested for the center, although it is not a dominant population. 
The contribution of an intermediate-age ($4 \sim 8$ Gyr) 
stellar population seems to be considerable at the galaxy center, 
but the contribution of such a population 
is small at the outskirts. 

\item
The evolution of Leo~II is suggested to be as follows.  
(1) Leo~II first began forming stars throughout the whole galaxy 
with a constant (inefficient) star-formation rate.  
(2) The star formation then began to cease in the outskirts and 
the star-forming region gradually shrank toward the center.  
(3) The star-forming activity had dropped to $\sim$ 0 by $\sim$ 4 Gyr ago 
except at the center, where a small population younger than 4 Gyr is found.

\end{itemize}


\acknowledgements

We thank the observatory staff of the Subaru Telescope 
for their excellent support. 
We are grateful to the anonymous referee 
for many valuable comments and suggestions 
which improve this paper very much.

\clearpage

\begin{table}
  \begin{center}
    \begin{tabular}{llll}\hline
      Band    & Date      & Exposure Time [sec]              & FWHM [arcsec]\\ \hline\hline
      $V$     & 2001.4.20 & 3000 (5$\times$600)              & 0.6 -- 0.8\\
              &           &  900 (5$\times$180)              & 0.6 -- 0.8\\
      $I$     & 2001.4.24 & 2400 (5$\times$240+4$\times$300) & 0.6 -- 0.8\\
              &           &  300 (5$\times$60)               & 0.6 -- 0.8\\ \hline
    \end{tabular}
    \caption{The log of the observation.}
    \label{tab:obs}
  \end{center}
\end{table}

\begin{table}
  \begin{center}
    \begin{tabular}{lllllll}\hline
      Area [arcmin] & $<r>$ [arcmin] & Bright RGB & Faint RGB & Blue HB & Red HB & SB [mag/arcsec$^{2}$] \\ \hline\hline
	$0.0 -- 0.5$   &  0.354  &  38 &  34 &   4 &  32 & 25.28 \\
	$0.5 -- 1.5$   &  1.120  & 203 & 227 &  13 & 184 & 25.60 \\
	$1.5 -- 2.5$   &  2.062  & 244 & 319 &  16 & 253 & 26.02 \\
	$2.5 -- 3.5$   &  3.041  & 220 & 307 &  20 & 193 & 26.77 \\
	$3.5 -- 4.5$   &  4.031  & 127 & 168 &  22 & 122 & 27.63 \\
	$4.5 -- 5.5$   &  5.025  &  62 &  98 &  12 &  52 & 28.69 \\
	$5.5 -- 6.5$   &  6.021  &  30 &  51 &   8 &  19 & 29.51 \\
	$6.5 -- 7.5$   &  7.018  &  18 &  20 &   5 &  25 & 30.29 \\
	$7.5 -- 8.5$   &  8.016  &   7 &  21 &   2 &  13 & 31.08 \\
	$8.5 -- 9.5$   &  9.014  &   5 &  16 &   1 &   6 & 31.76 \\
	$9.5 -- 10.5$  & 10.013  &   5 &  13 &   1 &   5 & 31.98 \\
	$10.5 -- 11.5$ & 11.011  &   5 &   9 &   0 &   6 & 32.45 \\
	$11.5 -- 12.5$ & 12.010  &   5 &  14 &   2 &   7 & 32.27 \\
	$12.5 -- 13.5$ & 12.923  &   4 &  10 &   1 &   7 & 32.38 \\
	$13.5 --$      & 14.215  &   4 &  18 &   2 &   9 & 32.97 \\ \hline
    \end{tabular}
    \caption{The number of stars in each area. 
	The integrated surface brightness for these components in V band 
	is listed in the right-most column. }
    \label{tab:radprof}
  \end{center}
\end{table}

\begin{table}
  \begin{center}
    \begin{tabular}{cccc}\hline
            & $f_{K,0}$     & $r_{c}$ [arcmin] & $r_{t}$ [arcmin]\\ \hline\hline
      Bright RGB &  77.6$\pm$6.5  & 2.28$\pm$0.30    & 9.22$\pm$0.53 \\
      Faint RGB  & 104.0$\pm$8.3  & 3.05$\pm$0.34    & 8.51$\pm$0.26 \\

      All RGB    & 183.4$\pm$13.2 & 2.76$\pm$0.28    & 8.63$\pm$0.26 \\

      Red  HB    & 96.2$\pm$11.2  & 3.24$\pm$0.48    & 6.99$\pm$0.22 \\
      Blue HB    &  5.3$\pm$0.8   & 4.05$\pm$0.78    & 10.78$\pm$0.78 \\ \hline
    \end{tabular}
    \caption{The best-fit parameters for King profile fitting. }
    \label{tab:king}
  \end{center}
\end{table}

\clearpage

\begin{figure}
  \caption{The color image of our survey area. 
           North is to the top and east is to the left. 
           Both width and height of the image are 26.67 arcmin. 
	  }
  \label{fig:leo2}
\end{figure}

\begin{figure}
  \plotone{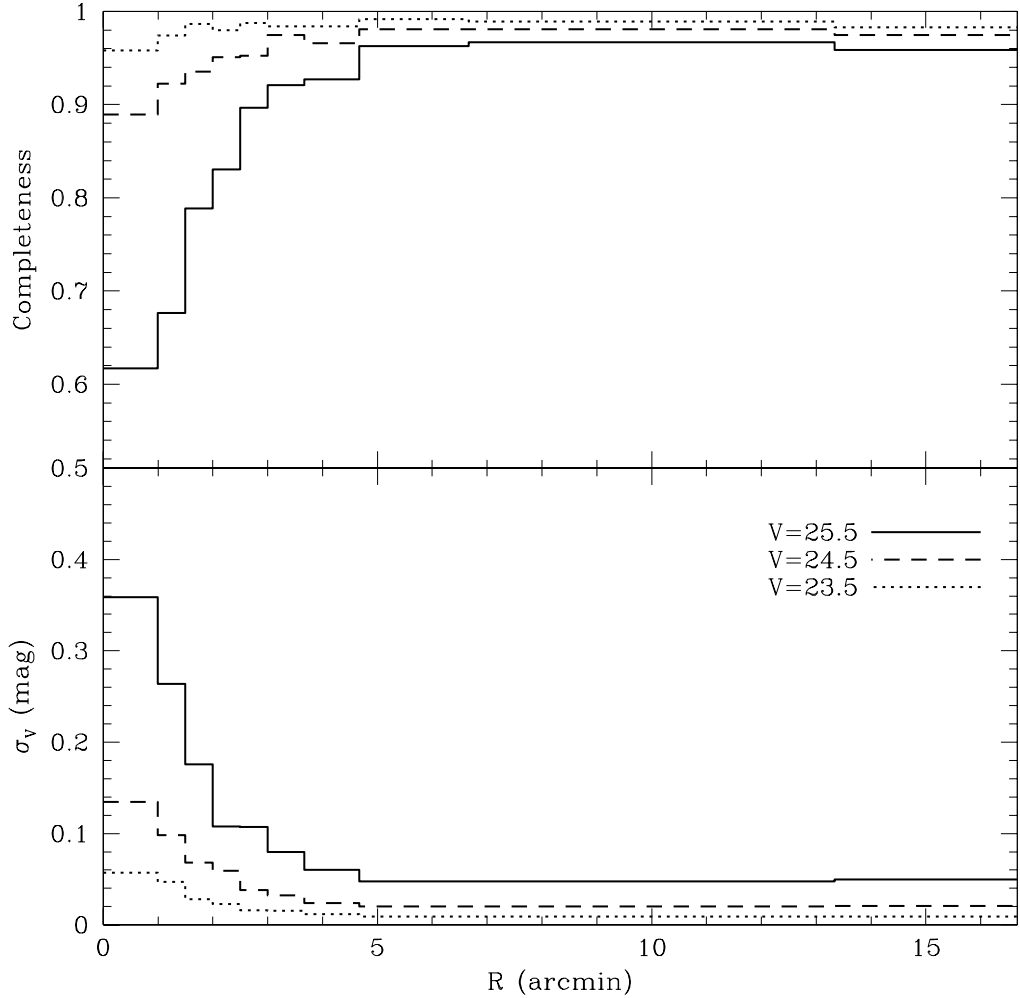}
  \caption{The detection completeness (top) and the magnitude errors (bottom) 
	   are plotted as a function of the distance from the galaxy center 
	   for different magnitude ($V=23.5, 24.5, 25.5$). 
           }
  \label{fig:complete}
\end{figure}%

\begin{figure}
  \caption{The color-magnitude diagram of stars in the central 
           $6.67 \times 6.67$ arcmin$^{2}$ field. 
           }
  \label{fig:cmd}
\end{figure}%

\begin{figure}
  \caption{(a) The criteria for RGB, blue and red HB selection  
           are overlaid on the color-magnitude diagram. 
           Typical photometric errors at the center ($\sim2.5$ arcmin) 
	   and the outskirts of the galaxy are indicated 
           as blue and red error bars at $V-I=1.4$. 
           (b) The detailed view of the RGB sequence.  
           The mean RGB sequence (Eq.\ref{eq:rgbseq}) is plotted in red 
           together with those of Galactic globular clusters 
           M~15, NGC~6397, M~2 and NGC~1851 (from left to right) in cyan. 
           (c) Detailed view of the SGB. 
           Padova isochrones for ages 5, 10, and 15 Gyr 
           (from top to bottom) of different metallicity population 
           (Z=0.0004 in green and Z=0.001 in magenta)
           are overlaid. 
           }
  \label{fig:cmfea}
\end{figure}%

\begin{figure}
  \plotone{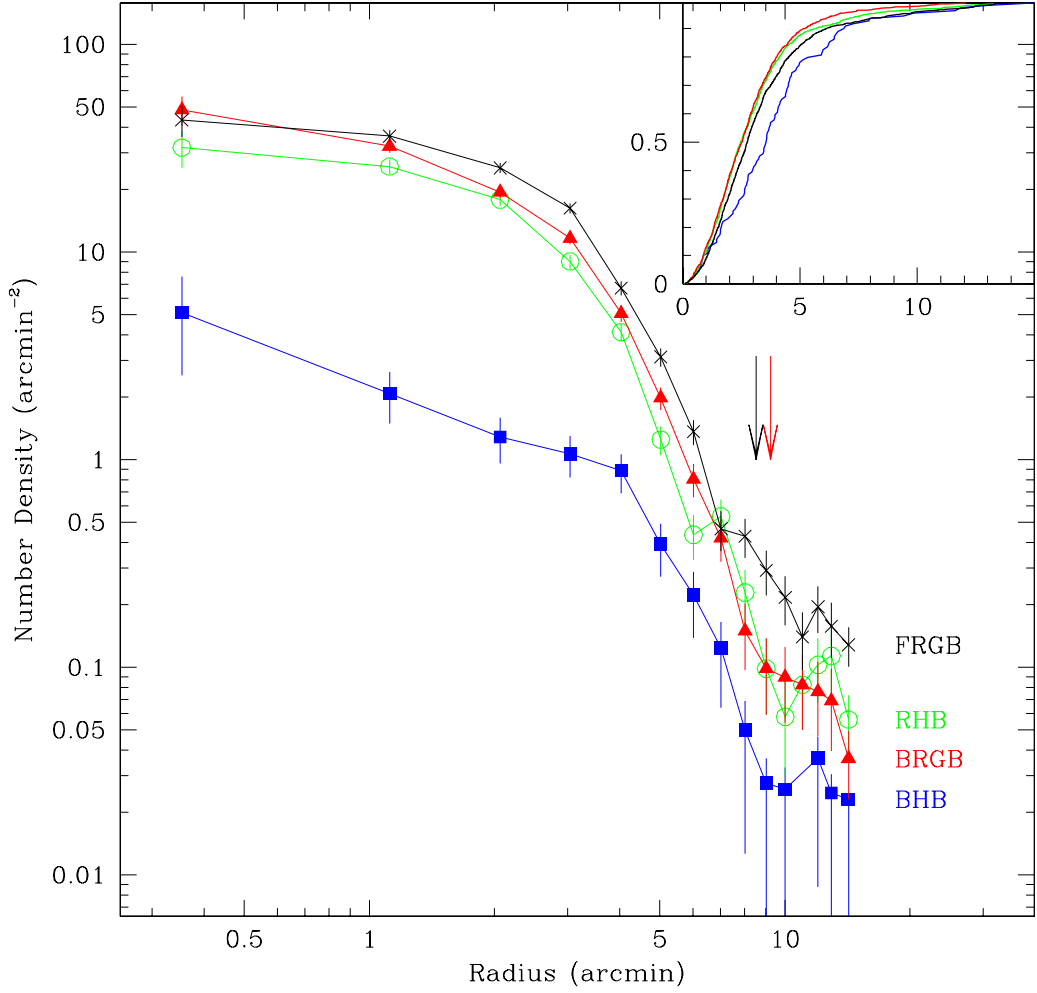}
  \caption{The radial profile of each stellar component. 
	   The red, black, blue and green lines represent 
	   the radial profiles of bright RGB, faint RGB, blue HB, and 
	   red HB stars, respectively. 
           The error bars are estimated on the basis of Poisson statistics. 
	   Two arrows indicate the tidal radii calculated for 
	   bright RGB (red) and faint RGB (black), respectively. 
           The inset shows 
           the cumulative number fraction of each stellar component
	   as a function of the radius 
	   in the same colors as described above. 
           }
  \label{fig:radprof}
\end{figure}%

\begin{figure}
  \caption{Smoothed surface brightness map of RGB and HB stars. 
           Contours correspond roughly to 
	   26.5, 27.5, 28.3, 29.0, 30.0 mag/arcsec$^{2}$ 
	   from the center. 
           }
  \label{fig:densmap}
\end{figure}%

\begin{figure}
  \plotone{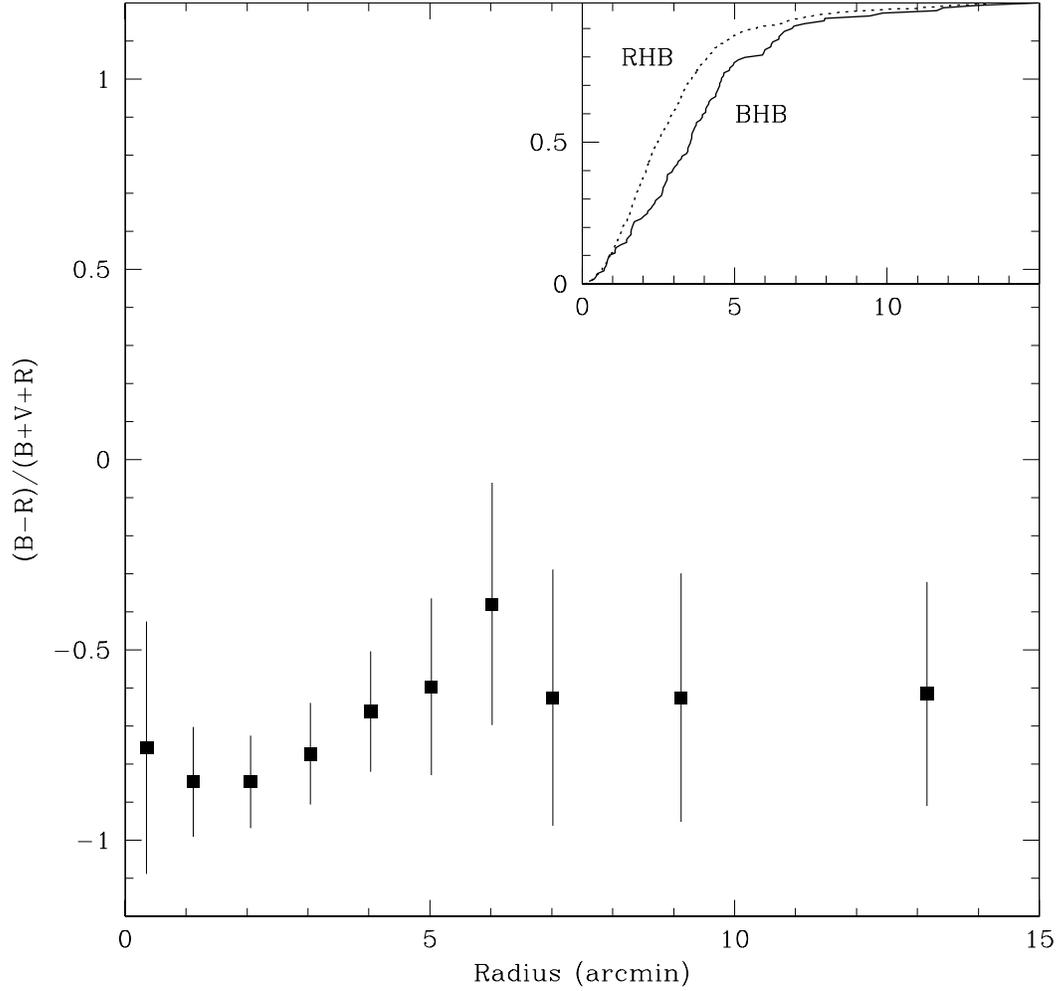}
  \caption{The HB morphology index $(B-R)/(B+V+R)$
           plotted as a function of the radius. 
           $B$, $R$ and $V$ indicate numbers of blue, red HB stars, and 
           those stars found at the RR Lyr instability strip, respectively. 
           The error bars are estimated based on the Poisson statistics. 
           The inset shows 
           the cumulative number fraction of blue (solid) 
           and red (dashed) HB stars as a function of the radius. 
           }
  \label{fig:hbmorph}
\end{figure}%

\begin{figure}
  \plotone{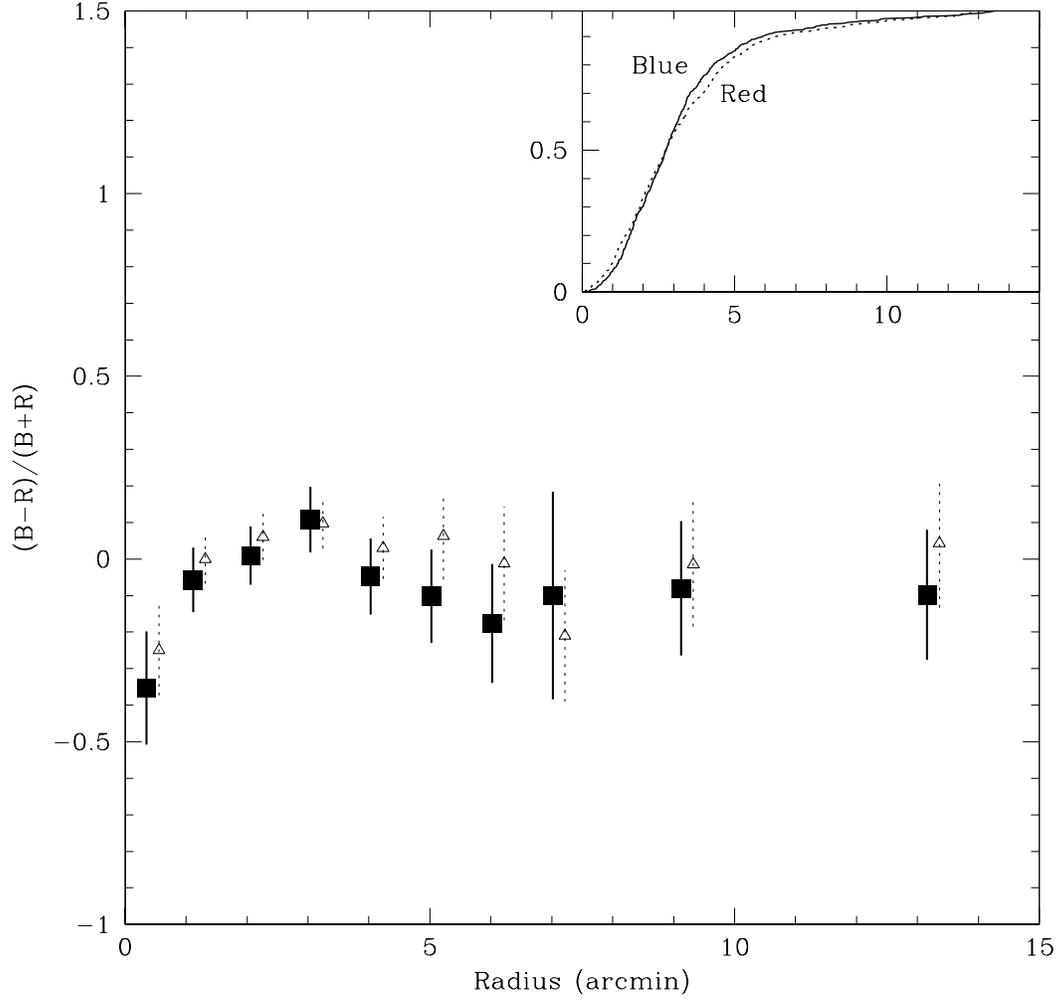}
  \caption{The RGB color index $(B-R)/(B+R)$
           plotted as a function of the radius. 
           $B$ and $R$ indicate numbers of stars 
           that deviate less than 0.075 mag bluer and redder 
           from the mean RGB sequence, respectively. 
           The indices derived from whole RGB stars ($19<V<23.5$) 
           and faint RGB stars ($22.18<V<23.5$)
           are plotted as open triangles and filled squares. 
           The error bars are estimated based on the Poisson statistics. 
           The inset shows 
           the cumulative number fraction of blue (solid) 
           and red (dashed) faint RGB stars as a function of the radius. 
           }
  \label{fig:rgbmorph}
\end{figure}%

\begin{figure}
  \plotone{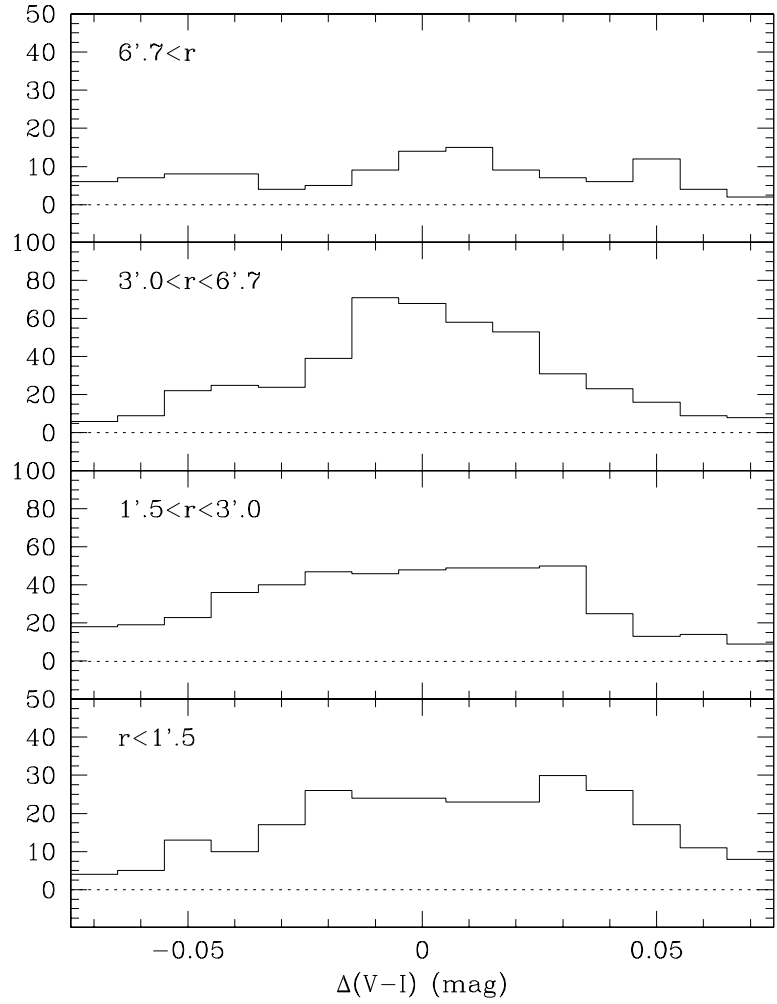}
  \caption{The color distribution of faint RGB stars 
           around the mean RGB sequence in different annuli 
           ($r<1'.5, 1'.5<r<3'.0, 3'.0<r<6'.7$, and $6'.7<r$). 
           }
  \label{fig:frgbchist}
\end{figure}%

\begin{figure}
  \plotone{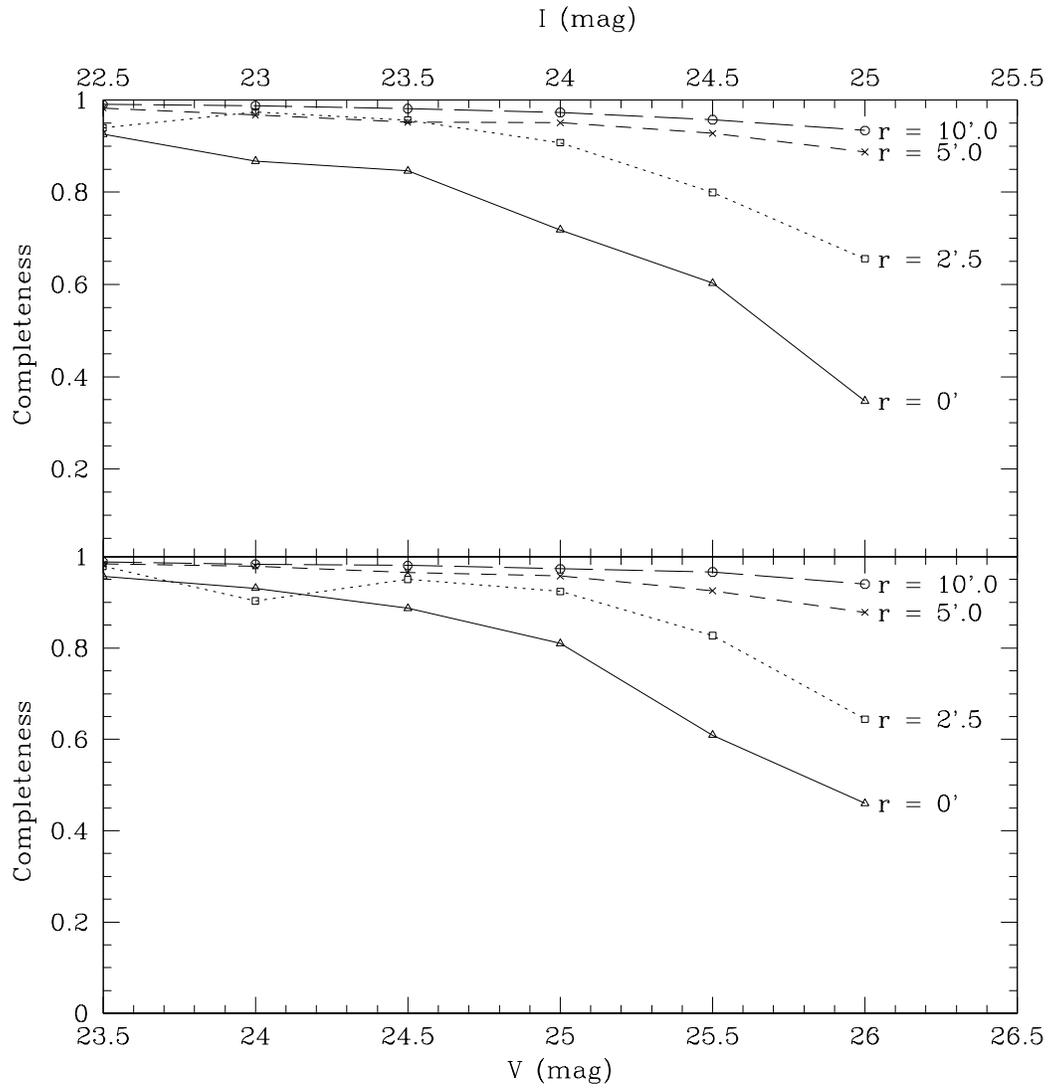}
  \caption{Detection completeness as a function of 
	magnitude in $V$ (bottom) and $I$ (top) bands for different radii 
	($r$ = 0, 2.5, 5.0, 10.0 arcmin). 
           }
  \label{fig:comp_mag}
\end{figure}%

\begin{figure}
  \plotone{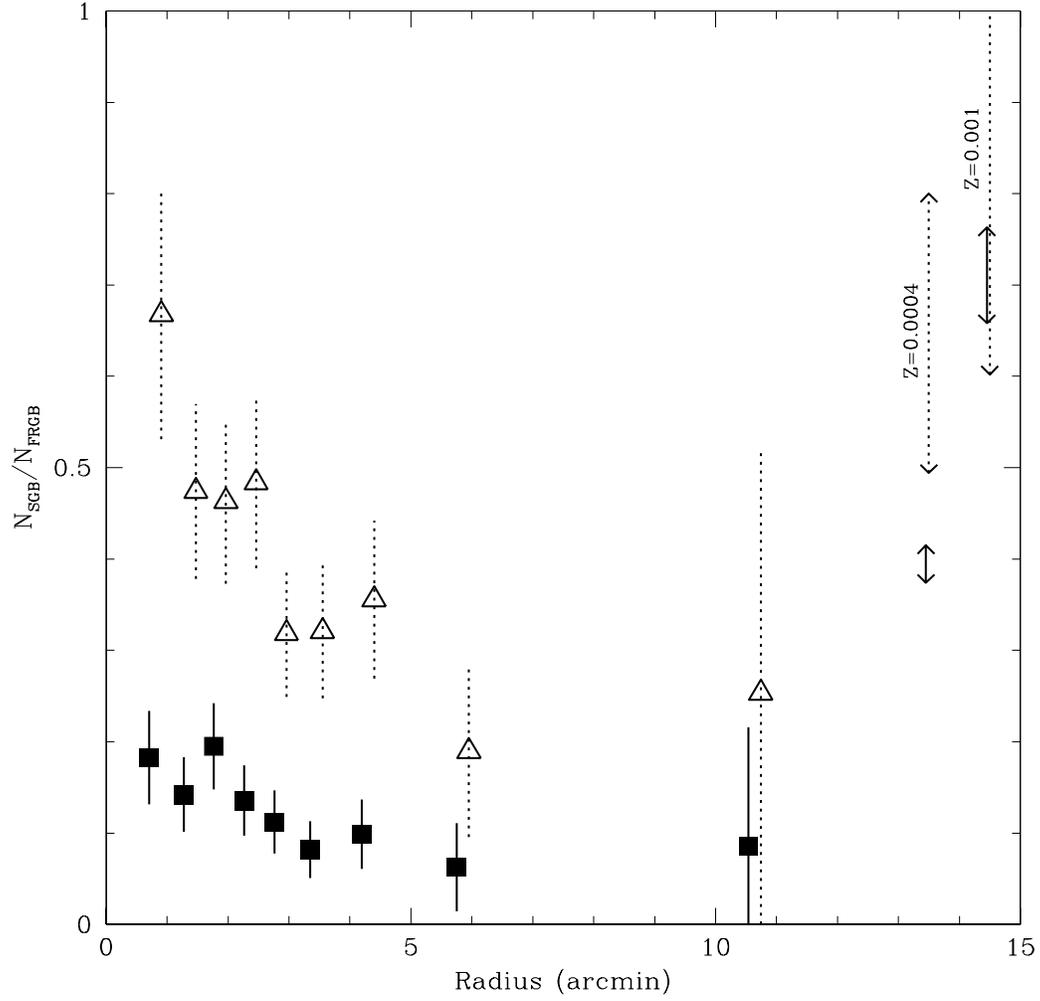}
  \caption{The number ratio of SGB to faint RGB stars 
	   plotted as a function of the radius. 
	   Filled squares and open triangles represent 
	   the number ratio for bright ($23.5<V<24.0$) and intermediate 
	   ($24.0<V<24.5$) SGB stars, respectively. 
	   The error bars are estimated on the basis of Poisson statistics. 
	   The solid and dotted arrows at $r \sim 14$ represent 
	   the calculated number ratios (see text) 
	   for bright and intermediate SGB stars, respectively, 
	   of different metallicities. 
           }
  \label{fig:sgfrgbratio}
\end{figure}%

\begin{figure}
  \plotone{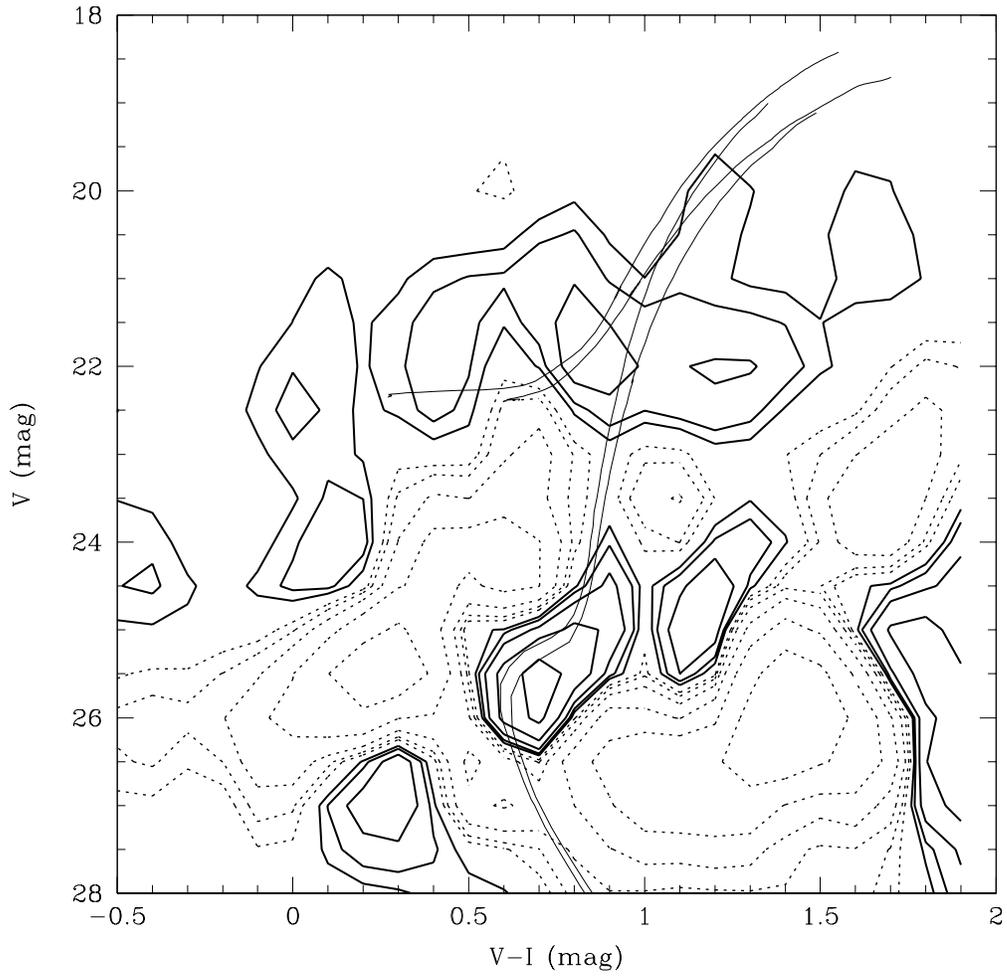}
  \caption{Field subtracted Hess diagram for the knot. 
	   The solid contours represent 1, 2, 4, 8, 16, 32 stars 
	   per $\Delta(V-I)=0.1$ and $\Delta V=0.5$ bin. 
	   The dotted contours represent -1, -2, -4, -8, -16, -32 stars, 
	   indicating that field contamination is oversubtracted. 
	   Two isochrones (Z=0.0004 and Z=0.001 with age of 15 Gyr) 
	   are overlaid for the guidance. 
           }
  \label{fig:hess}
\end{figure}%

\end{document}